\documentclass{sigchi-ext}
\usepackage[T1]{fontenc}
\usepackage{textcomp}
\usepackage[scaled=.92]{helvet} 
\usepackage{graphicx} 
\usepackage{balance}  
\usepackage{booktabs} 
\usepackage{ccicons}  
\usepackage{ragged2e} 

\usepackage{wasysym} 
\usepackage{fixltx2e} 

\def\plaintitle{Exploratory Study of Young Children's Social Media Needs and Requirements} 
\def\emptyauthor{}
\def\plainkeywords{Children; Social Network; Design Workshop; Participatory.}

\title{Exploratory Study of Young Children's Social Media Needs and Requirements}

\numberofauthors{3}
\author{
  \alignauthor{
    \textbf{Di ``Chelsea'' Sun}\\
    \affaddr{Human-Computer Interaction Group} \\
    \affaddr{University of California, Merced} \\
    \affaddr{Merced, CA 95343, USA} \\
    \email{dsun23@ucmerced.edu}}
    \vfil\alignauthor{
    \textbf{Vaishnavi Melkote}\\
    \affaddr{Human-Computer Interaction Group} \\
    \affaddr{University of California, Merced} \\
    \affaddr{Merced, CA 95343, USA} \\
    \email{vmelkote2@ucmerced.edu}}
    \vfil\alignauthor{
    \textbf{Ahmed Sabbir Arif}\\
    \affaddr{Human-Computer Interaction Group} \\
    \affaddr{University of California, Merced} \\
    \affaddr{Merced, CA 95343, USA} \\
    \email{asarif@ucmerced.edu} \\
    \url{https://www.asarif.com}}
}

\definecolor{linkColor}{RGB}{6,125,233}
\hypersetup{%
  pdftitle={\plaintitle},
  pdfauthor={\emptyauthor},
  pdfkeywords={\plainkeywords},
  bookmarksnumbered,
  pdfstartview={FitH},
  colorlinks,
  citecolor=black,
  filecolor=black,
  linkcolor=black,
  urlcolor=linkColor,
  breaklinks=true,
}

\copyrightinfo{\scriptsize Permission to make digital or hard copies of part or all of this work for personal or classroom use is granted without fee provided that copies are not made or distributed for profit or commercial advantage and that copies bear this notice and the full citation on the first page. Copyrights for third-party components of this work must be honored. For all other uses, contact the owner/author(s). \\
{\emph{Interaction Design and Children (IDC '20 Extended Abstracts), June 21--24, 2020, London, United Kingdom.}} \\
\copyright~2020 Copyright is held by the author/owner(s). \\
ACM ISBN 978-1-4503-8020-1/20/06. \\
http://dx.doi.org/10.1145/3397617.3397836}

\begin{document}

\maketitle

\RaggedRight{} 

\begin{abstract}
As social media are becoming increasingly popular among young children, it is important to explore this population's needs and requirements from these platforms. As a first step to this, we conducted an exploratory design workshop with children aged between ten and eleven years to find out about their social media needs and requirements. Through an analysis of the paper prototypes solicited from the workshop, here we discuss the social media features that are the most desired by this population.
\end{abstract}




\section{Introduction}
For better or worse, social media are becoming increasingly popular among young children. Nowadays, children younger than twelve years are frequently accessing various social media platforms. There has also been in a rise of social media apps for mobile devices targeted at children aged between 3 and 13. Many have investigated the effects of social media on young children and proposed ways to mitigate the negative effects by limiting and monitoring children's social media access and usage. Others have made recommendations on how to design age-appropriate social media and social media content. Most of these works, however, investigated children's social media activities from an adult's point of view, without much consideration for what children find interesting and useful. Hence, the recommendations made in these works are not always effective in practice, instead can cause distress among children, and create conflict and tension between them and their parents. We argue that an understanding of young children's needs and requirements from social media is necessary to tackle the negative effects of social media on this population. This exploratory work takes an aim at it by conducting a design workshop with children aged between 10 and 11 years.

\begin{figure}[t]
  \centering
  \includegraphics[width=\columnwidth]{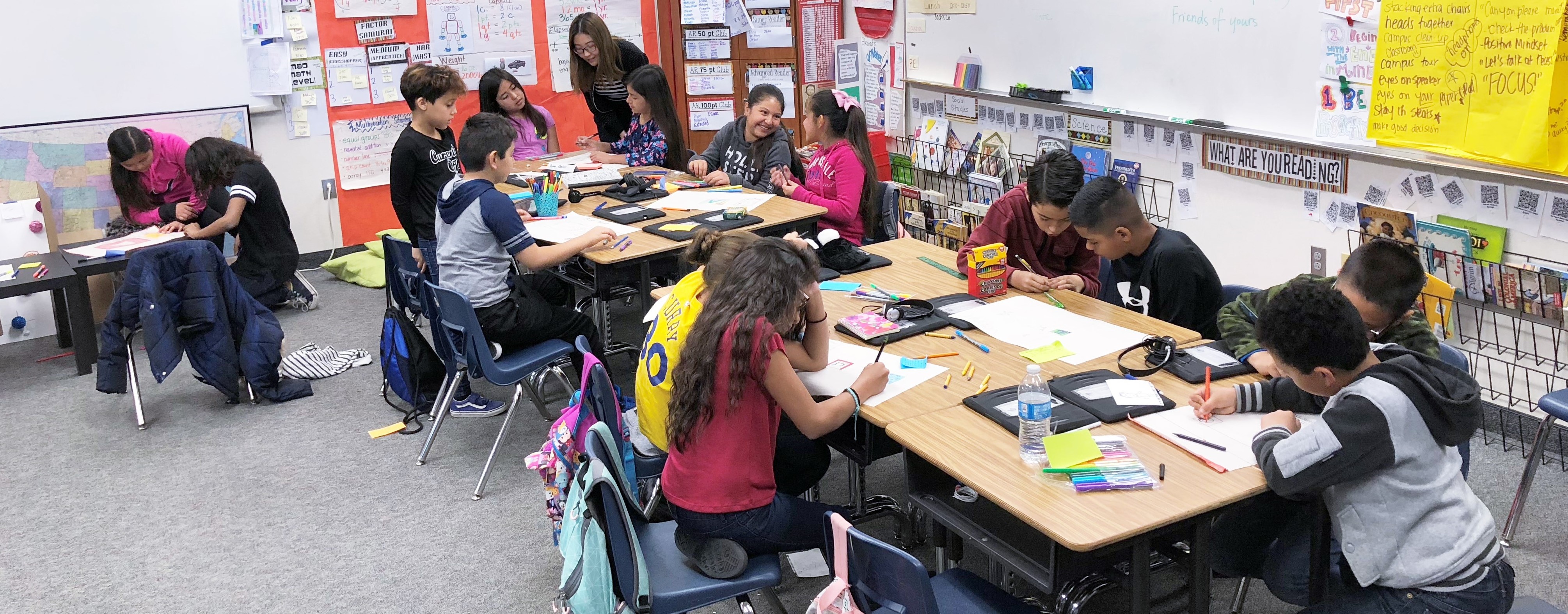}
  \caption{Children participating in the design workshop.}~\label{fig:workshop}
\end{figure}

\section{Related Work}
Many have explored how and why teenagers and young adults use social media, and how the platforms influence their lives and education, but not many have focused on children aged between 11 and 12. Livingstone \cite{livingstone_developing_2014} proposed a social developmental pathway where \textit{``children are first recipients, then participants, and finally actors in their social media worlds``}. Informed by media literacy research, she reported that children aged between 9 and 11 years are ``recipient'' who  question what is real or fake on social media, children between 11 and 13 are ``participants'' who are more interested in what is fun, even if it is transgressive or fake, and children between 14 and 16 are ``actors'' who refocus on what is valuable for them.

Most research involving children aged between 11 and 12 caution about the risks of using social media. Based on a literature review, Richards et al. \cite{richards_impact_2015} concluded that social media affect children's mental health, specifically in the areas of self-esteem, body image, and well-being, primarily due to cyberbullying. Hamm et al. \cite{hamm_prevalence_2015} also reported potential associations between cyberbullying and depression, self-harm, and suicidality. Hinduja and Patchin \cite{hinduja_bullying_2010} reported that middle-schoolers exposed to cyberbullying are more likely to have suicidal thoughts and attempt suicide. They found out non-white children tend to have higher levels suicidal ideation than white children. Ybarra and Mitchell \cite{ybarra_how_2008} reported that 15\% children aged between 10 and 15 encounter unwanted sexual solicitation on social media. Palfrey et al. \cite{federal2009empowering} emphasized on the potential long-term effects of sharing too much details on social media. Young children unaware of privacy issues often post inappropriate messages, pictures, and videos, which may affect their personal, academic, and professional life in the future. In addition, Winpenny et al. \cite{winpenny_exposure_2014} cautioned parents and policy makers on adolescent exposure to online marketing of addictive substances, particularly alcohol. O'Keeffe et al. \cite{okeeffe_impact_2011} and Chassiakos et al. \cite{chassiakos_children_2016} identified several risks of young children using social media, particularly negative effects on health, sleep, attention, and learning; cyberbullying and online harassment; a higher incidence of obesity and depression \cite{jelenchick_facebook_2013}; exposure to inaccurate, inappropriate, and unsafe content and contacts; and compromised privacy and confidentiality. As a countermeasure, the American Academy of Pediatrics \cite{media_children_2013} encouraged establishing a \textit{family home use plan} for digital media. They discouraged media exposure to children younger than 2 years and recommended limiting young children's media access and usage within two hours daily, as well as monitoring their online activities.

Ito et al. \cite{ito_living_2009}, in contrast, encouraged teenagers and young children's engagement with social media with the argument that it provides \textit{``a degree of freedom and autonomy for youth that is less apparent in a classroom setting''}. They found out that teenagers and young children are heavily engaged in friendship-driven online activities, where they extend existing friendships through social media. Some children also engage in interest-driven activities, where they explore interests and acquire information. They reported that children usually pick up new skills and different forms of technical and media literacy by participating in these activities. Chassiakos et al. \cite{chassiakos_children_2016} also reported a number of evidence-based benefits of young children using social media, particularly \textit{``early learning, exposure to new ideas and knowledge, increased opportunities for social contact and support, and new opportunities to access health promotion messages and information''}. Moser et al. \cite{moser_parents_2017} investigated early teens and their parents' preferences about the parents sharing information about their children on social media. Results revealed that \textit{``parents and children are in agreement in their perception of how often and how much information parents share about their children on social media"}, but children believe their parents should ask permission more often than parents think they should.

\begin{margintable}[-40pc]
  \begin{minipage}{\marginparwidth}
    \centering
    \begin{tabular}{r r l}
        {\small \textbf{Mobile Activity}}
      & {\small \textbf{N}}
      & {\small \textbf{\%}} \\
       \toprule
       Education & 14 & 100 \\
       Internet Search & 14 & 100 \\
       Social Media & 14 & 100 \\
       Video Streaming & 9 & 64 \\
       Online Chat & 8 & 57 \\
       Video Games & 6 & 43 \\
       Music Streaming & 4 & 29 \\
      \bottomrule
    \end{tabular}
    \caption{The most popular online activities on mobile devices. Note that percentages do not add up to 100\% as children performed more than one activities.}~\label{tab:activities}
  \end{minipage}
\end{margintable}

\begin{margintable}[-10pc]
  \begin{minipage}{\marginparwidth}
    \centering
    \begin{tabular}{r r l}
        {\small \textbf{Social Media}}
      & {\small \textbf{N}}
      & {\small \textbf{\%}} \\
      \toprule
      YouTube & 8 & 57 \\
      Snapchat & 4 & 29 \\
      Musical.ly & 4 & 29 \\
      Facebook & 2 & 14 \\
      Instagram & 2 & 14 \\
      Snapchat & 2 & 14 \\
      Twitter & 2 & 14 \\
      \bottomrule
    \end{tabular}
    \caption{About 93\% children had their own social media accounts. Note that percentages do not add up to 100\% since some children had multiple accounts.}~\label{tab:socialmedia}
  \end{minipage}
\end{margintable}

\section{Design Workshop}
We conducted a 3-hour workshop (including recess) at a 5th-grade classroom at Lorena Falasco Elementary School in Los Banos, California to learn about young children's social media access and usage, and their most desired and disliked features of social media. We handed out Informed Consent and Media Release forms to all students ahead of time to acquire permission from their parents to participate in the workshop. On the day of the workshop, an Assent Form was used to collect children's verbal consents. Participation occurred only when the parents and the child both agreed. 14 students took part in the workshop. Their age ranged from 10 to 11 years. 8 of them were female and 6 were male. They all were frequent users of mobile devices. The workshop was carried out in a classroom (Figure \ref{fig:workshop}). Children who did not receive permission from their parents were taken to the other side of the room for a different activity organized by the class teacher. The workshop included the following activities, with 15-minute recess in between.

\textbf{Focus Group (60 minutes)} The workshop started with a focus group to learn about children's social media access and usage. The researcher asked open-ended questions and recorded all responses. She made sure that children understood the questions and lead the conversation to restrict deviation from the topic. She asked follow-up questions for further clarification, when necessary.

\textbf{Creative Session (60 minutes)} In this session, children were instructed to write, draw, or create paper mock-ups of their most desired social media features. For this, we provided them with different types of papers (construction paper, color copier paper, plain white paper, sticky notes, index cards), a variety of writing utensils (pens, pencils, colored pencils, crayons, markers), scissors, regular glue, glue sticks, and tapes. A researcher walked around the room and spent time with all participants to get a grasp of their ideas (Figure \ref{fig:workshop}), however, refrained from showing enthusiasm or disagreement to avoid the possibility of influencing their process. Children were encouraged to share and discuss their ideas with classmates, thus helped each other fine-tuning their ideas (a collective design process).

\textbf{Debrief Session (30 minutes)} The workshop ended with children discussed their ideas with the class, the teacher, and the researcher. The researcher asked follow-up questions to get an understanding of all design aspects.

\section{Usage Behaviors}
All participants had access to desktop and laptop computers, smartphones, and tablets both at home and school. They mainly used tablets for in-class activities, and sometimes desktop computers located in a designated room at the school. They used smartphones and tablets to perform a variety of tasks (Table \ref{tab:activities}). All participants used social media on mobile devices using either their own accounts or the accounts of their parents or guardians. About 93\% children had at least one social media account, some had accounts for multiple platforms (Table \ref{tab:socialmedia}). This is interesting since most of these platforms require users to be at least 13 years of age to sign up for an account. However, children informed us that they are allowed to use social media only when they abide by the rules set by their parents. Some common rules were: a specific number of hours on social media per day, only talking and interacting with family members and friends, staying offline when eating, no swearing, and conforming with parents when adding new friends.

\begin{marginfigure}[-25pc]
  \begin{minipage}{\marginparwidth}
    \centering
    \includegraphics[width=0.9\marginparwidth]{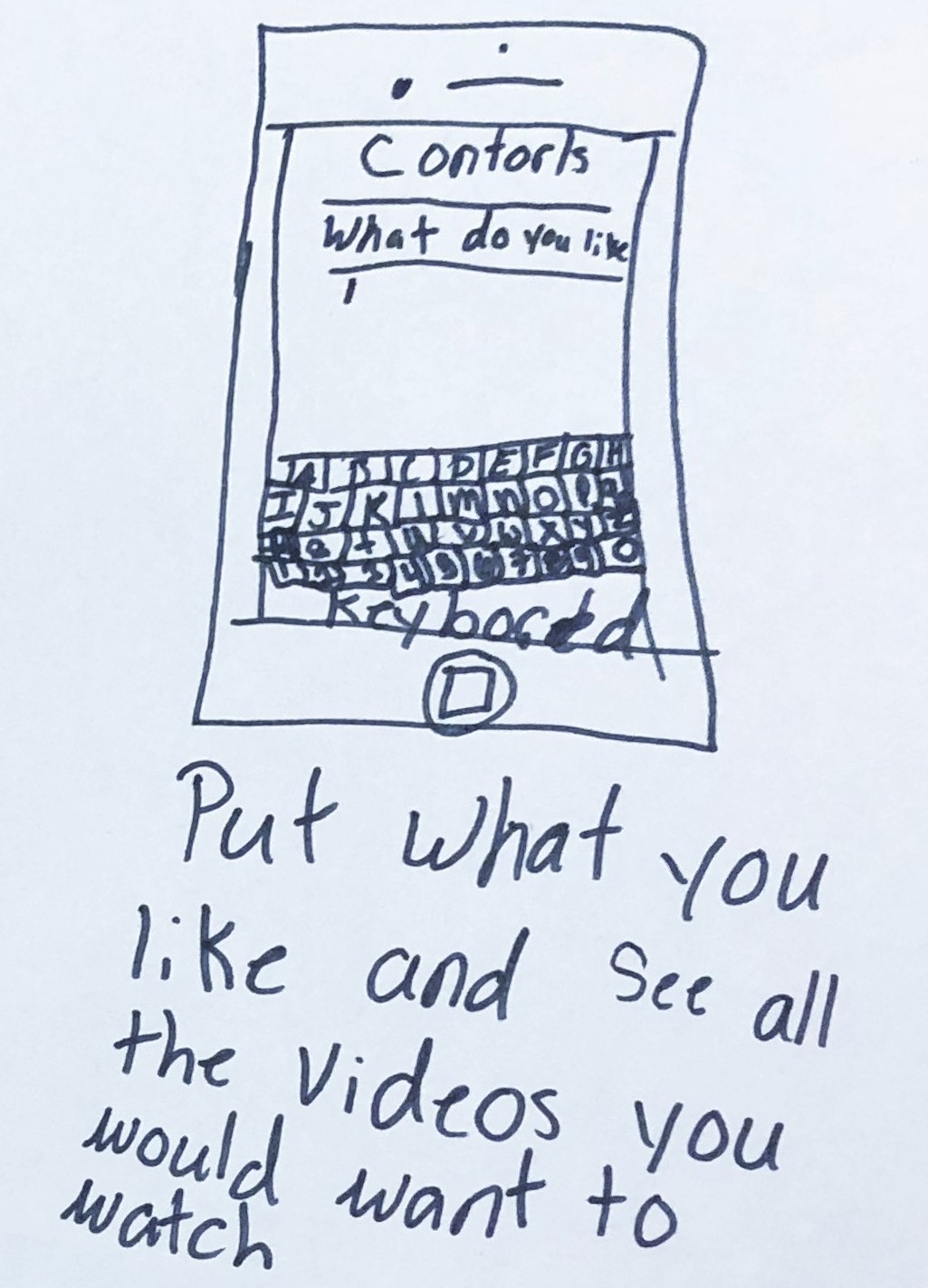}
    \caption{A design expressing the desire for better, personalized recommender systems for video suggestion.}~\label{fig:youtube-rec}
  \end{minipage}
\end{marginfigure}

\begin{marginfigure}[0pc]
  \begin{minipage}{\marginparwidth}
    \centering
    \includegraphics[width=0.9\marginparwidth]{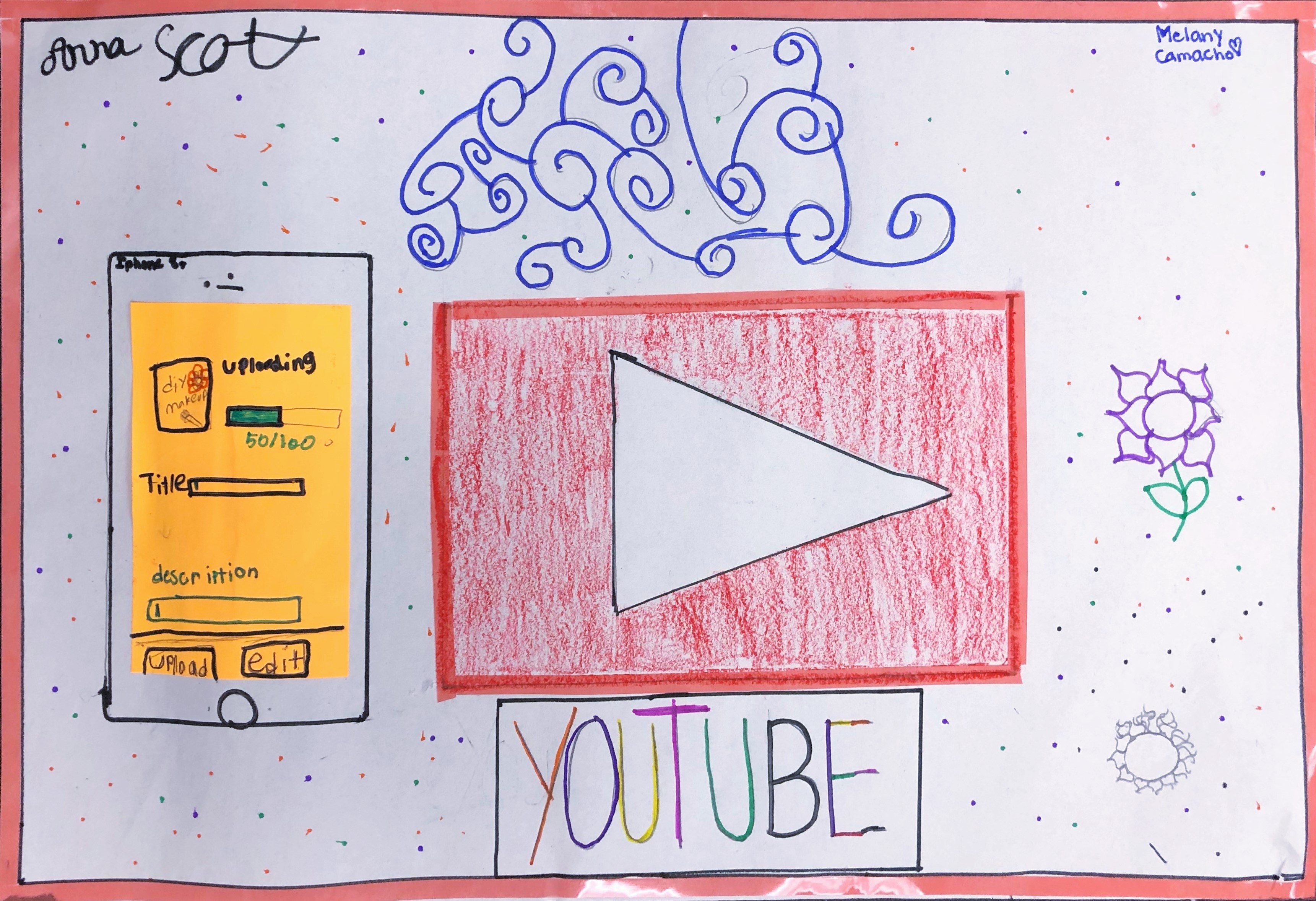}
    \caption{A design highlighting the need for easier video editing and annotation methods.}~\label{fig:youtube-anno}
  \end{minipage}
\end{marginfigure}

Children used social media mainly to communicate with family members. Most of them used these platforms to share pictures and recordings of video games, and to chat with family members (particularly aunts and grandparents). They never used social media to make new friends, and rarely to connect with classmates. However, some children expressed interest in connecting with unknown child users with shared interests. Investigation revealed that children preferred using different social media for different online activities. For example, they preferred Google for educational purposes, i.e., to find reading material or to get help with assignments. YouTube was the preferred platform for watching funny videos and videos of music, lip-sync, product reviews, and experienced gamers playing video games. Some children also liked watching live-streamed videos of family members and gaming events. Although watching videos was a vital part of children's social media activities, they almost never contributed any content. Only a few uploaded videos to share their experience with family members' or to react to their videos. Children informed us that they only created and shared content on platforms that are user-friendly. They stressed how difficult it is to produce and share good-quality content on most social media. When they ran into any issues with these platforms, they sought assistance from their older siblings, parents, or someone who had more experience with technology. Interestingly, children preferred making new friends and connecting with each other in-person rather then online because they enjoyed observing each other's reactions and hearing each other's voices in conversations. Most of them disliked online messaging as they frequently made spelling or grammar mistakes, and the auto-correct feature of most virtual keyboards changed the word they meant to type \cite{arif_learning_2016}.

\section{Needs and Requirements}
We carefully studied the design outcomes of the workshop to identify social media features that are the most desired by the children, summarized below in no particular order.

\textbf{Acquiring and Producing Multimedia.} Children expressed a strong desire for acquiring multimedia content on social media, especially videos. They complained that popular social media platforms are cluttered with irrelevant and dull videos, which makes finding the videos they like difficult. They wanted an efficient personalized recommender system that can suggest videos based on user-selected categories and the videos they have rated, shared, commented on, and the content creators they follow on social media (Figure \ref{fig:youtube-rec}). Children also expressed interest in producing multimedia content, particularly images and videos. They criticized the current image and video production capabilities of popular social media platforms, thus wanted these to be simpler and more efficient. They also desired the ability to edit pictures and videos, and annotate with texts, shapes, images, and emojis (Figure \ref{fig:youtube-anno}). They felt that providing easy-to-follow instructions on how to capture images and videos, how to edit and annotate, and how to upload and stream multimedia could enhance the quality and quantity of the content created by child users. In keeping with the current trends, some children wanted an easy mechanism for converting pictures and videos to animations (GIFs) and memes. They also wanted to keep updated on how well their content is doing in terms of likes, shares, and comments, preferably via notifications.

\begin{marginfigure}[-10pc]
  \begin{minipage}{\marginparwidth}
    \centering
    \includegraphics[width=0.9\marginparwidth]{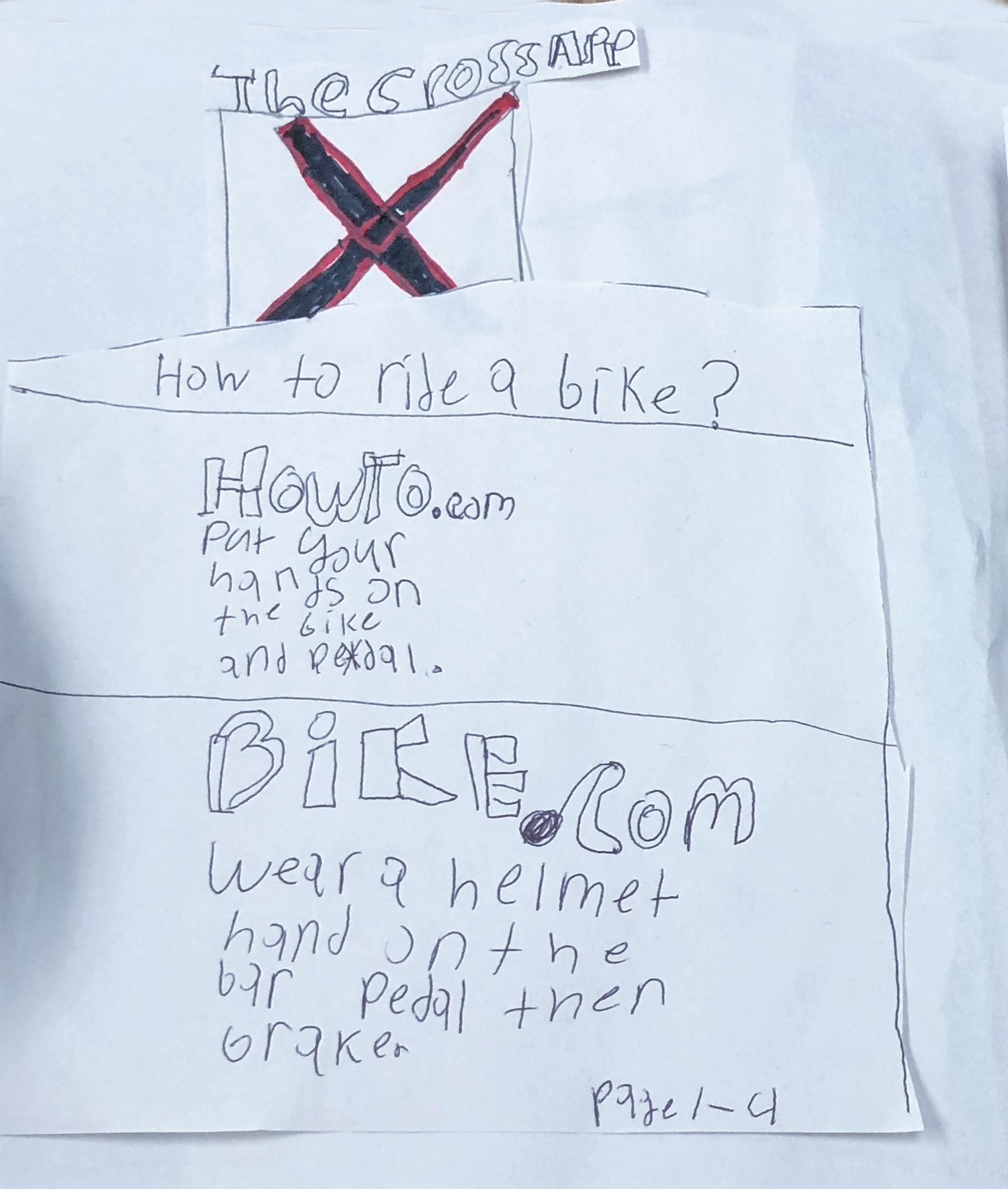}
    \caption{A design expressing the desire for easy-to-follow tutorials on social media to acquire new skills.}~\label{fig:tutorial}
  \end{minipage}
\end{marginfigure}

\begin{marginfigure}[-0pc]
  \begin{minipage}{\marginparwidth}
    \centering
    \includegraphics[width=0.9\marginparwidth]{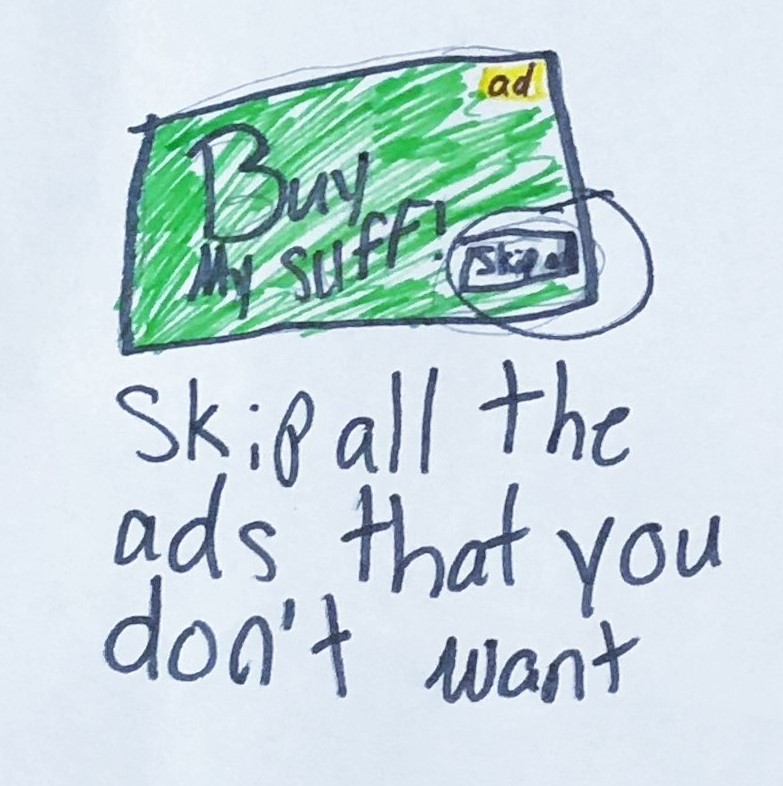}
    \caption{Children wanted better controls for the types of ads they see on social media.}~\label{fig:ads}
  \end{minipage}
\end{marginfigure}

\begin{marginfigure}[1pc]
  \begin{minipage}{\marginparwidth}
    \centering
    \includegraphics[width=0.9\marginparwidth]{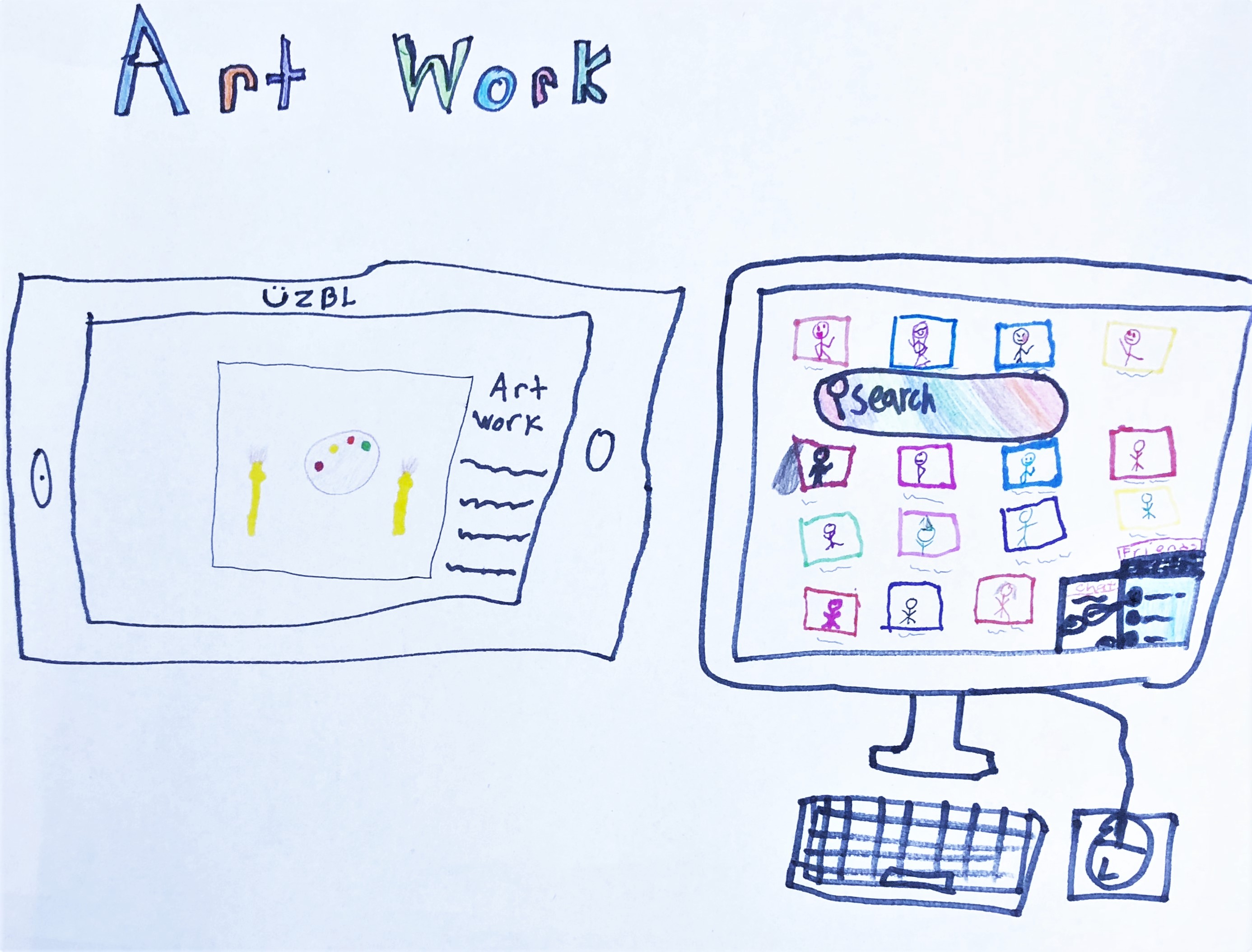}
    \caption{Children expressed interest in social media features that can foster their creativity.}~\label{fig:coloring}
  \end{minipage}
\end{marginfigure}

\textbf{Learning and Creativity.} Children were interested in learning new skills on social media. They wanted these platforms to make an effort to connect children with shared interests so that they can inspire and learn from each other. They also wanted easy-to-follow tutorials and video instructions on how to master new skills on social media, such as how to play a musical instrument or ride a bicycle (Figure \ref{fig:tutorial}). Some were keen on features that can foster their creativity, particularly digital painting, coloring, and creative writing (Figure \ref{fig:coloring}). They also wanted better networking abilities with like-minded children and step-by-step tutorials. Some suggested a reward-based approach to encourage learning, such as earning points when they achieve specific goals. They felt this will keep them interested in the activities and promote healthy competitions among their peers.

\textbf{Digital Advertisement.} Children expressed a strong distaste for the digital advertisements they see on social media (Figure \ref{fig:ads}). They found most of these ads boring, irrelevant, annoying, or inappropriate, thus wanted options for avoid seeing the ads they do not like. It is interesting that children did not demand ad-free platforms (could be because they were not even aware that it was a possibility), instead wanted a better control over the types of ads they see.

\textbf{Design and Usability.} Children preferred colorful user interfaces as they felt vibrant colors make interfaces attractive. They also wanted simpler, more natural user interfaces, stressing how difficult it is for them to use current social media platforms. They also wanted interaction methods that do not require performing many actions, for example a ``single-click'' option for recording and uploading videos.

\textbf{Online Gaming.} Some children wanted a wider selection of freely available multiplayer games on social media. Some of them also wanted the option to share games they own with their peers to be able to play together, an option currently unavailable on social media.

\section{Conclusion}
We conducted a design workshop to explore young children's social media usage behaviors and most desired social media features. The findings of this exploratory work provide an understanding of this population's social media interactions and expectations, and most importantly directions for future research. It is important that social media aimed at young children are designed keeping the population's needs and desires in mind to providing them with a pleasant, productive, and safe social media experience that do not put them at risk. In the future, we will conduct follow-up studies with a larger sample to further investigate some of the patterns identified in this work. We will also extend our investigation to parents and teachers to find out their take on children's responses.

We thank the Merced County Office of Education and Lorena Falasco Elementary School for their help and support.

\balance{}

\bibliographystyle{SIGCHI-Reference-Format}
\bibliography{extended-abstract}


\begin{thebibliography}{00}


\ifx \showCODEN    \undefined \def \showCODEN     #1{\unskip}     \fi
\ifx \showDOI      \undefined \def \showDOI       #1{{\tt DOI:}\penalty0{#1}\ }
  \fi
\ifx \showISBNx    \undefined \def \showISBNx     #1{\unskip}     \fi
\ifx \showISBNxiii \undefined \def \showISBNxiii  #1{\unskip}     \fi
\ifx \showISSN     \undefined \def \showISSN      #1{\unskip}     \fi
\ifx \showLCCN     \undefined \def \showLCCN      #1{\unskip}     \fi
\ifx \shownote     \undefined \def \shownote      #1{#1}          \fi
\ifx \showarticletitle \undefined \def \showarticletitle #1{#1}   \fi
\ifx \showURL      \undefined \def \showURL       #1{#1}          \fi

\bibitem{arif_learning_2016}
{Ahmed~Sabbir Arif}, {Cristina Sylla}, {and} {Ali Mazalek}. 2016.
\newblock \showarticletitle{Learning {New} {Words} and {Spelling} with
  {Autocorrections}}. In {\em Proceedings of the 2016 {ACM} {International}
  {Conference} on {Interactive} {Surfaces} and {Spaces}} {\em ({ISS} '16)}.
  ACM, New York, NY, USA, 409--414.
\newblock
\showISBNx{978-1-4503-4248-3}


\bibitem{chassiakos_children_2016}
{Yolanda (Linda)~Reid Chassiakos}, {Jenny Radesky}, {Dimitri Christakis},
  {Megan~A. Moreno}, {Corinn Cross}, {and} {Council on Communications~And
  Media}. 2016.
\newblock \showarticletitle{Children and {Adolescents} and {Digital} {Media}}.
\newblock {\em Pediatrics\/} {138}, 5 (Nov. 2016).
\newblock
\showISSN{0031-4005, 1098-4275}


\bibitem{hamm_prevalence_2015}
{Michele~P. Hamm}, {Amanda~S. Newton}, {Annabritt Chisholm}, {Jocelyn Shulhan},
  {Andrea Milne}, {Purnima Sundar}, {Heather Ennis}, {Shannon~D. Scott}, {and}
  {Lisa Hartling}. 2015.
\newblock \showarticletitle{Prevalence and {Effect} of {Cyberbullying} on
  {Children} and {Young} {People}: {A} {Scoping} {Review} of {Social} {Media}
  {Studies}}.
\newblock {\em JAMA Pediatrics\/} {169}, 8 (Aug. 2015), 770--777.
\newblock
\showISSN{2168-6203}


\bibitem{hinduja_bullying_2010}
{Sameer Hinduja} {and} {Justin~W. Patchin}. 2010.
\newblock \showarticletitle{Bullying, {Cyberbullying}, and {Suicide}}.
\newblock {\em Archives of Suicide Research\/} {14}, 3 (July 2010), 206--221.
\newblock
\showISSN{1381-1118}


\bibitem{ito_living_2009}
{Mizuko Ito}. 2009.
\newblock {\em Living and {Learning} with {New} {Media}: {Summary} of
  {Findings} from the {Digital} {Youth} {Project}}.
\newblock MIT Press.
\newblock
\showISBNx{978-0-262-25827-2}


\bibitem{jelenchick_facebook_2013}
{Lauren~A. Jelenchick}, {Jens~C. Eickhoff}, {and} {Megan~A. Moreno}. 2013.
\newblock \showarticletitle{“{Facebook} {Depression}?” {Social}
  {Networking} {Site} {Use} and {Depression} in {Older} {Adolescents}}.
\newblock {\em Journal of Adolescent Health\/} {52}, 1 (Jan. 2013), 128--130.
\newblock
\showISSN{1054-139X}


\bibitem{livingstone_developing_2014}
{Sonia Livingstone}. 2014.
\newblock \showarticletitle{Developing {Social} {Media} {Literacy}: {How}
  {Children} {Learn} to {Interpret} {Risky} {Opportunities} on {Social}
  {Network} {Sites}}.
\newblock {\em Communications\/} {39}, 3 (2014), 283--303.
\newblock
\showISSN{1613-4087}


\bibitem{media_children_2013}
{Council on Communications~And Media}. 2013.
\newblock \showarticletitle{Children, {Adolescents}, and the {Media}}.
\newblock {\em Pediatrics\/} {132}, 5 (Nov. 2013), 958--961.
\newblock
\showISSN{0031-4005, 1098-4275}


\bibitem{moser_parents_2017}
{Carol Moser}, {Tianying Chen}, {and} {Sarita~Y. Schoenebeck}. 2017.
\newblock \showarticletitle{Parents' and {Children}'s {Preferences} about
  {Parents} {Sharing} about {Children} on {Social} {Media}}. In {\em
  Proceedings of the 2017 {CHI} {Conference} on {Human} {Factors} in
  {Computing} {Systems}} {\em ({CHI} '17)}. Association for Computing
  Machinery, Denver, Colorado, USA, 5221--5225.
\newblock
\showISBNx{978-1-4503-4655-9}


\bibitem{okeeffe_impact_2011}
{Gwenn~Schurgin O'Keeffe}, {Kathleen Clarke-Pearson}, {and} {Council on
  Communications~and Media}. 2011.
\newblock \showarticletitle{The {Impact} of {Social} {Media} on {Children},
  {Adolescents}, and {Families}}.
\newblock {\em Pediatrics\/} {127}, 4 (April 2011), 800--804.
\newblock
\showISSN{0031-4005, 1098-4275}


\bibitem{federal2009empowering}
{John Palfrey}, {Urs Gasser}, {and} {Danah Boyd}. 2010.
\newblock \showarticletitle{Empowering parents and protecting children in an
  evolving media landscape}.
\newblock {\em Response to FCC Notice of Inquiry\/} (2010), 09--94.
\newblock


\bibitem{richards_impact_2015}
{Deborah Richards}, {Patrina~HY Caldwell}, {and} {Henry Go}. 2015.
\newblock \showarticletitle{Impact of social media on the health of children
  and young people}.
\newblock {\em Journal of Paediatrics and Child Health\/} {51}, 12 (2015),
  1152--1157.
\newblock
\showISSN{1440-1754}


\bibitem{winpenny_exposure_2014}
{Eleanor~M. Winpenny}, {Theresa~M. Marteau}, {and} {Ellen Nolte}. 2014.
\newblock \showarticletitle{Exposure of {Children} and {Adolescents} to
  {Alcohol} {Marketing} on {Social} {Media} {Websites}}.
\newblock {\em Alcohol and Alcoholism\/} {49}, 2 (March 2014), 154--159.
\newblock
\showISSN{0735-0414}


\bibitem{ybarra_how_2008}
{Michele~L. Ybarra} {and} {Kimberly~J. Mitchell}. 2008.
\newblock \showarticletitle{How {Risky} {Are} {Social} {Networking} {Sites}?
  {A} {Comparison} of {Places} {Online} {Where} {Youth} {Sexual} {Solicitation}
  and {Harassment} {Occurs}}.
\newblock {\em Pediatrics\/} {121}, 2 (Feb. 2008), e350--e357.
\newblock
\showISSN{0031-4005, 1098-4275}


\end{thebibliography}

\end{document}